\documentclass[aps,twocolumn,pra,tightenlines,floatfix,showpacs]{revtex4-1}

\usepackage[dvips]{graphicx}
\usepackage[english]{babel}
\usepackage{amsmath}
\usepackage{amssymb}
\usepackage{times}
\def\be{\begin{equation}}
\def\ee{\end{equation}}
\def\bea{\begin{eqnarray}}
\def\eea{\end{eqnarray}}

\begin{document}

\title{Reentrant superfluidity and pair density wave in single component dipolar Fermi gases}

\author{Yanming Che}
\affiliation{\vspace*{-2.ex}Zhejiang Institute of Modern Physics and Department of Physics,
Zhejiang University, Hangzhou, Zhejiang 310027, China and Synergetic Innovation Center of Quantum Information and
  Quantum Physics, Hefei, Anhui 230026, China} 

\author{Jibiao Wang}
\affiliation{\vspace*{-2.ex}Zhejiang Institute of Modern Physics and Department of Physics,
Zhejiang University, Hangzhou, Zhejiang 310027, China and Synergetic Innovation Center of Quantum Information and
  Quantum Physics, Hefei, Anhui 230026, China} 

\author{Qijin Chen}
\email[\vspace*{-2.5ex}Corresponding author: ]{qchen@zju.edu.cn}
\affiliation{\vspace*{-2.ex}Zhejiang Institute of Modern Physics and Department of Physics,
Zhejiang University, Hangzhou, Zhejiang 310027, China and Synergetic Innovation Center of Quantum Information and
  Quantum Physics, Hefei, Anhui 230026, China} 

\pacs{03.75.Ss,67.85.Lm,74.20.Rp,74.25.Dw \hfill Journal ref: Phys. Rev. A \textbf{93}, 063611 (2016)}

\date{Mar 16, 2015; \today}

\begin{abstract}
  We study the superfluidity of single component dipolar Fermi gases
  in three dimensions using a pairing fluctuation theory, within the
  context of BCS-BEC crossover.  The transition temperature $T_{c}$
  for the dominant $p_z$ wave superfluidity exhibits a remarkable
  re-entrant behavior as a function of the pairing strength induced by
  the dipole-dipole interaction (DDI), which leads to an anisotropic
  pair dispersion. The anisotropy and the long range nature of the DDI
  cause $T_c$ to vanish for a narrow range of intermediate interaction
  strengths, where a pair density wave emerges as the ground
  state. The superfluid density and thermodynamics below $T_{c}$,
  along with the density profiles in a harmonic trap, are investigated
  as well. Implications for experiments are discussed.
\end{abstract}

\maketitle

\section{Introduction}

\vspace*{-1.5ex}
Recent experimental realization of quantum degenerate Fermi gases of
magnetic atoms \cite{Mingwu,Chromium-dipolar,GrimmPRL.112.010404} and
the rapid progress toward creating degenerate polar molecules
\cite{JinScience,Zwierlein,ZwierleinPRL.114.205302} have opened a new
frontier for exploring novel phases of quantum gases, where
dipole-dipole interaction (DDI) plays a central role.  A lot of
attentions have been paid to unconventional $p$-wave superfluids
\cite{You,Baranov,VincentLiu-PRL,WuHirsch,LiWu} in three dimensions (3D) and
topological superfluids \cite{Cooper} in two dimensions (2D). The
latter has been associated with Majorana fermions and can be used for
topological quantum computation \cite{Qi}. Such exotic superfluid
phases emerge from the long-range DDI with a strong anisotropy, which
differs from the widely studied contact potential in dilute atomic
gases. Moreover, the relative DDI strength can be tuned by changing
the fermion number density $n$ (or Fermi wavevector $k_F$) and, in the
case of polar molecules \cite{JinNature}, by varying an external
electric field strength.

Of particular interest is the intermediate pairing strength regime,
where complex physics beyond the weak coupling BCS theory arises and
the superfluid transition temperature $T_{c}$ is relatively high,
making it more practical to access the superfluid phase
experimentally. For a contact potential, the entire BCS--Bose-Einstein
condensation (BEC) crossover from weak to strong coupling regimes has
been studied intensively in two-component Fermi gases of $^6$Li or
$^{40}$K. In contrast, such a crossover in dipolar Fermi gases, where
richer physics may arise, is yet to be explored. Existing theoretical
studies in this aspect mostly focus on the ground state, based on mean
field treatments \cite{You,Baranov,Zhai,Yi}, which are inadequate in
addressing moderate and strong coupling regimes at finite temperature.

In this paper, we address the superfluidity and pairing phenomena of
\emph{single component} dipolar Fermi gases in 3D, with an emphasis on
the finite temperature and interaction effects.
Built on previous work \cite{ChenPRL,ChenReports} that has been
applied successfully to address various BCS-BEC crossover phenomena in
two-component Fermi gases with a contact interaction
\cite{ChenReports,FrontPhys}, here we construct a similar pairing
fluctuation theory for the superfluidity of fully polarized
\emph{one-component} dipolar fermions (in the $\hat{z}$ direction), in
which thermally excited pairs naturally give rise to a pseudogap in
the fermion excitation spectrum. We find that (i) the
DDI leads dominantly to a $p_z$-wave superfluid, and the superfluid
$T_c$ curve exhibits a re-entrant behavior as a function of the DDI
strength; in the intermediate regime of the BCS-BEC crossover, $T_c$
vanishes and the ground state becomes a pair density wave (PDW),
similar to the PDW state studied in underdoped high $T_c$
superconductors \cite{TesanovicCPCDW,PDWZhang}. (ii) In the fermionic
regime, the temperature dependence of superfluid density and low $T$
thermodynamic quantities exhibit power laws, as expected but in stark
contrast to the contact interaction case \cite{GrimmNature}.  (iii)
Within a local density approximation (LDA), the density profile in an
isotropic harmonic trap exhibits a similar qualitative behavior to its
$s$-wave counterpart, despite the different pairing symmetry and the
anisotropic pair mass.

The emergence of the PDW state originates from the long range nature
of the DDI, which essentially put the system in the high density
regime. The $p_z$-wave symmetry leads further to a non-local effect
\cite{Kosztin_nonlocal} and hence a diverging coherence length in the
nodal $xy$-plane, \emph{which makes it difficult for the pairs to move
  in the $\hat{z}$ direction, without heavily colliding with each
  other.}  At certain intermediate pairing strength, the interaction
energy between pairs may dominate the kinetic energy, in favor of
forming a Wigner-like crystal in the $\hat{z}$ direction. This PDW
state may exhibit behaviors of a Bose metal
\cite{BoseMetal1,*BoseMetal2,Yang_BoseMetal}, with a Bose ``surface''
for pair excitations at a finite pair momentum $q_z$ (with
$q_x=q_y=0$). The two dimensionality of the pair dispersion in the
remaining $xy$-plane destroys possible long range superfluid order,
leading to a metallic ground state with a density wave of Cooper pairs
in the $\hat{z}$ direction.

\vspace*{-1.5ex}
\section{Theoretical Formalism}
\vspace*{-1.5ex}
We consider an ultracold gas of one-component dipolar fermions of mass
$m$ in unit volume, with dipole moment $\mathbf{d}
=d\mathbf{\hat{z}}$, fully polarized in the $\hat{z}$ direction.  We
follow the pairing fluctuation theory as described in
Ref.~\cite{ChenPRL}, with fermion energy $\xi_{\mathbf{k}}={\bf
  k}^2/(2m)-\mu$ measured with respect to the chemical potential
$\mu$. (we take $\hbar=k_{B}=1$, as usual).  We shall write the
pairing interaction $V_{\mathbf{k},\mathbf{k'}}$ into an effective
separable form \cite{NSR}, i.e., $V_{\mathbf{k},\mathbf{k'}}=g
\varphi_{\mathbf{k}} \varphi_{\mathbf{k'}}^{*}$, where $g$ is the
pairing strength, $\varphi_{\mathbf{k}}$ is the symmetry factor with
an odd parity and will be determined by the DDI.

\begin{figure}
\centerline{\includegraphics[height=0.8in,width=3.in,clip]{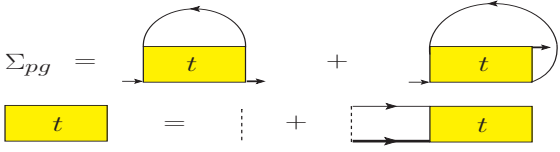}}
\caption{(Color online) Feynman diagrams for the pairing fluctuation
  self-energy $\Sigma_{pg}$ and $T$-matrix $t$. The thin solid, thick
  solid and dashed lines represent the bare propagator $G_0$, dressed
  propagator $G$ and DDI, respectively.\vspace*{-1.ex}}
\label{fig:SigmaPG}
\end{figure}

Following previous work \cite{Kadanoff&Martin,ChenPRL,ChenPhD}, the
fermion self-energy comes from particle-particle scattering, which
leads to both an order parameter (below $T_c$) and a pseudogap.
Noncondensed pairs are treated on an equal footing with single
particle propagators. In contrast to the $s$-wave singlet pairing case
\cite{ChenPRL}, an extra exchange diagram has now been retained in the
self energy, as shown in Fig.~\ref{fig:SigmaPG}. Besides the pairing
symmetry, \emph{this exchange diagram is a major difference between
  singlet and triplet pairing}.  Therefore, we obtain the fermion
self-energy from noncondensed pairs
\begin{eqnarray}
  \Sigma_{pg}(K)&=& \Sigma_{pg}^\text{direct}(K) + \Sigma_{pg}^\text{exchange}(K)\nonumber\\
  &=& \sum_{Q \neq 0} t(Q)G_0(Q-K)\varphi_{\mathbf{k}-\mathbf{q}/2}
  \varphi^{*}_{\mathbf{k}-\mathbf{q}/2}\nonumber\\
  &-& \sum_{Q \neq 0} t(Q)G_0(Q-K)\varphi_{{\mathbf{k}}-\mathbf{q}/2}\varphi^{*}_{{3}\mathbf{q}/2-\mathbf{k}}\,,\vspace*{-1.ex}
\label{eq: SigmaPG}
\end{eqnarray}
where $t(Q)=1/[g^{-1}+\chi(Q)]$, with $\chi(Q)=\sum_K G(K)
G_0(Q-K)|\varphi_{{\mathbf{k}}-{\mathbf{q}}/2}|^2$, and $G_0$ ($G$)
the bare (full) fermion Green's function. 
Below $T_c$, the condensate self-energy is\vspace*{-1.5ex}
\begin{equation}
\Sigma_{sc}(K)=-\Delta^2_{sc}G_0(-K)|\varphi_{\mathbf{k}}|^2,\vspace*{-0.5ex}
\end{equation}
 as in
BCS theory, with the superfluid order parameter $\Delta_{sc}$.
As in Ref.~\cite{ChenPRL}, we use a four vector notation, $K\equiv (i
\omega_{n}, \bf k)$, $Q \equiv (i \Omega_{l}, \bf q)$, $\sum_Q\equiv
T\sum_{l}\sum_\mathbf{q}$, etc., with $\omega_{n}$ ($\Omega_{l}$)
being odd (even) Matsubara frequencies.  Here
$\varphi^{*}_{\mathbf{k}}$ is the hermitian conjugate of
$\varphi_{\mathbf{k}}$.

We emphasize that the derivation of this theory is independent
  of the concrete form of the pairing interaction, namely, it is not
  essential whether the interaction is $s$-wave, $p$-wave or $d$-wave,
  short range or long range, provided that one can assume a separable
potential in the scattering $T$-matrix \cite{ChenPhD}. In fact, the
original zero temperature BCS-BEC crossover by Leggett was done with
$p$-wave pairing \cite{Leggett}.

Due to the anisotropy of the DDI, the pair dispersion acquires an
anisotropy as well, in contrast to the short-range contact potential
case in a two-component Fermi gas.  Namely, the finite $\bf q$ pair
propagator $t_{pg}(Q)$ can be expanded as \vspace*{-1ex}
\begin{equation}
  t_{pg}^{-1}(Q)=Z(i
  \Omega_l-\Omega_{\mathbf{q}}+\mu_{pair}+i\Gamma_{\Omega,\mathbf{q}})\,,\vspace*{-0.5ex}
\end{equation}
with an effective pair dispersion $\Omega_{\bf q}={{\bf
    q}_{\tiny{\perp}}^{2}}/(2M_{\tiny{\perp}}^{*}) +
q_{z}^{2}/(2M_{z}^{*})$ and an effective pair chemical potential
$\mu_{pair}$. Here the inverse residue $Z$ and the (anisotropic)
effective pair mass $M_{\tiny{\perp}}^{*}=M_x^* = M_y^*$ and
$M_{z}^{*}$ can be determined in the process of Taylor expansion, as
usual. 
Following Ref.~\cite{ChenPRL}, $\Sigma_{pg}$ can be approximated as
$\Sigma_{pg}(K)\approx -\Delta^2_{pg}G_0(-K)|\varphi_{\mathbf{k}}|^2$.
With the odd parity $\varphi_{-\mathbf{k}}=-\varphi_{\mathbf{k}}$,
here we have defined the pseudogap $\Delta_{pg}$ as\vspace*{-1ex}
\begin{equation}
 \Delta^2_{pg}=-2\sum_Q t_{pg}(Q) \approx 2Z^{-1}\sum_\mathbf{q}b(\Omega_\mathbf{q})\,,\vspace*{-0.5ex}
 \label{eq:PG}
\end{equation}
where $b(x)$ is the Bose distribution function. This leads to the BCS
form of the total self-energy, 
\begin{equation}
  \Sigma(K)=
  \Sigma_{sc}(K)+\Sigma_{pg}(K) =
  -\Delta^2G_0(-K)|\varphi_{\mathbf{k}}|^2,
\end{equation}
with a total excitation gap
$\Delta=\sqrt{\Delta^2_{sc}+\Delta^2_{pg}}$.

As in Ref.~\cite{ChenPRL}, 
 from the Thouless criteria, $t^{-1}(0,{\bf{0}}) = 0$,
we have the gap equation\vspace*{-1ex}
\begin{equation}
  1+g\sum_{\mathbf{k}} \frac{1-
    2f(E_{\mathbf{k}})}{2E_{\mathbf{k}}}|\varphi_{\mathbf{k}}|^2 = 0\,,\vspace*{-0.5ex}
\label{eq: Gap Equation}
\end{equation}
and the fermion number equation
\begin{equation}
  n=\sum_{K}  G(K)= \sum_{\mathbf{k}}\left[
\frac{1}{2}\left(1-\frac{\xi_{\bf{k}}} {E_{\bf{k}}}\right)
+  \frac{\xi_{\mathbf{k}}}{E_{\mathbf{k}}}\,f(E_{\mathbf{k}})\right]\,,
\label{eq:number}
\end{equation}
where $E_{\mathbf{k}}=\sqrt{\xi_{\mathbf{k}}^2 +
  \Delta^{2}|\varphi_{\mathbf{k}}|^{2}}$ is the Bogoliubov
quasiparticle dispersion and $f(x)$ the Fermi distribution function.

Now we determine the symmetry factor $\varphi_{\mathbf{k}}$ from the DDI, 
\begin{equation}
  V_{d}(\mathbf{r})
 = d^2 \frac{1-3\cos^{2}{\theta_{\mathbf{r}}}}{r^{3}} = V(r)Y_{2,0}(\theta_{\mathbf{r}},\phi_{\mathbf{r}}),
\label{DDI}
\end{equation}
%
where the radial part $V(r)=-\sqrt{16\pi/5}\,{d^2}/{r^{3}}$, 
and the angular part $Y_{2,0}(\theta_{\mathbf{r}},\phi_{\mathbf{r}})$ is the
spherical harmonic $Y_{lm_l}(\hat{\bf{r}})$, with
$\theta_{\mathbf{r}}$ and $\phi_{\mathbf{r}}$ the polar and azimuthal angles of $\bf{r}$. 
So the DDI breaks SO(3) symmetry and mixes different partial waves.
Expanding $V_{\bf{k},\bf{k}'}$ in terms of partial waves, we have
$V_{\bf{k},\bf{k}'}=
\sum_{ll'}\sum_{m_{l}m_{l'}}g_{m_{l}m_{l'}}^{ll'}(k,k')Y_{lm_l}(\hat{\bf{k}})Y^{*}_{l'm_{l'}}(\hat{\bf{k}'})$,
with $g_{m_{l}m_{l'}}^{ll'}(k,k')= (-1)^{\frac{3l+l'}{2}}{16 \pi^2}
w_{l,l'}(k,k')\langle lm_{l}|Y_{20}|l'm_{l'} \rangle $ and
$w_{l,l'}(k,k')=\int_0^\infty r^2\mathrm{d} r
j_{l}(kr)V(r)j_{l'}(k'r)$, where $j_{l}(kr)$ is the spherical Bessel
function. For a single component Fermi gas, only odd $l$ and $l'$ are
allowed, with $l'=l, l\pm2$. The $r^{-3}$ dependence of the DDI leads
to a $k$-independent $w_{l,l'}(k,k)$. Detailed analyses show that the
dominant \textit{attractive} channel in $V_{\bf{k},\bf{k}}$ is $l=1$,
$m_l=0$, i.e., the $p_z$ wave, where $g_{00}^{11}(k,k)<0$ is the
leading order term, with $g_{00}^{33}(k,k)\approx0.1g_{00}^{11}(k,k)$
being the next leading order term. The leading hybridization terms
with $l=1$, $l'=3$ are repulsive.  Therefore
here 
we concentrate on the $p_{z}$-wave channel.

\begin{figure}
\centerline{\includegraphics[height=1.4in,width=3.4in,clip]{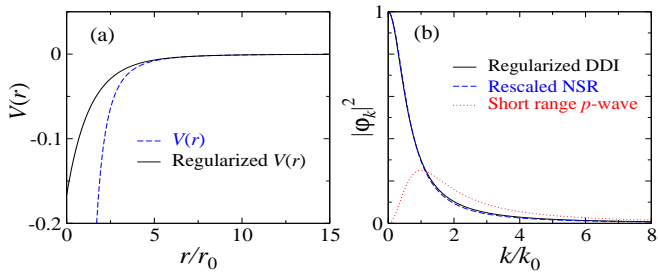}}
\caption{(Color online) (a) Radial part, $V(r)$, and regularized
  $V(r)$ of the DDI, in units of $\sqrt{16\pi/5}\,d^{2}$. (b) $k$
  dependence of $|\varphi_{\bf{k}}|^{2}$ calculated from the
  regularized DDI (black solid line) and
  $|\varphi_{\bf{k}}|_\text{NSR}^{2}$ (blue dashed). For comparison,
  the radial part of a short-range interaction induced $p$-wave
  $|\varphi_{\bf{k}}|_p^{2}$, which scales as $k^2$ in the low energy
  limit, is plotted as well (red dotted).}
\label{fig:DDI}
\end{figure}

To remove the ultraviolet divergence in the momentum integral of the
gap equation, caused by the $k$ independence of $w_{l,l'}(k,k)$, we
regularize the DDI by multiplying a convergence factor
$F({r}/{r_{0}})$, where $r_{0}$ is the typical radius beyond which the
DDI becomes dominant \cite{NoteonRegularization}. We choose
$F(x)=1-e^{-x}(1+x+{x^{2}}/{2})$, similar to that used in
Ref.~\cite{Zhai} but here the regularized DDI approaches a finite
value as $r \rightarrow 0$, as shown in Fig.~\ref{fig:DDI}(a).  This
leads to a modified $p_z$-wave symmetry factor\vspace*{-1.5ex}
\begin{equation}
 \varphi_{\mathbf{k}}^{2}
 = \frac{1}{2 \eta ^{2}}\left[1-\frac{\ln(1+4 \eta ^{2})}{4 \eta ^{2}}\right]
 \cos^{2}{\theta_{\mathbf{k}}}\,, \vspace*{-1.ex}
\label{symmetry factor}
\end{equation}
where $\varphi_{\mathbf{k}}$ is real, with $\eta=k/k_{0}=kr_{0}$, and
$\theta_{\mathbf{k}}$ the polar angle of $\bf{k}$. Interestingly, the
$k$ dependence of this $\varphi_{\mathbf{k}}$ is quantitatively very
close to a rescaled $s$-wave Lorentzian symmetry factor used in
Ref.~\cite{NSR}, \vspace*{-1.5ex}
\begin{equation}
  \left.\varphi_{\bf{k}}^{2}\right|_\text{NSR}^{}=\frac{1}{1+(1.55
    k/k_{0})^{2}},\vspace*{-1.ex}
\end{equation}
 as shown in Fig.~\ref{fig:DDI}(b). For comparison,
we also plot the $k$ dependence of a typical $p$-wave symmetry factor,
$|\varphi_{k}|^{2}_{p}=\dfrac{(k/k_{0})^{2}} {[1+(k/k_{0})^{2}]^{2}}$,
induced by a short-range interaction \cite{TLHo,Iskin,Ohashi}, for
which the partial wave scattering amplitude $f^{l}_{k} \sim V_{kk}
\sim |\varphi_{\bf{k}}|^{2} \sim a_{l} k^{2l}$ as $k \rightarrow 0$ so
that for $l=1$, $a_{1}$ is the scattering volume. \emph{In contrast, the
behavior of the $p_z$-wave scattering amplitude of the DDI is very
similar to the short range $s$-wave case, giving rise to a
well-defined scattering length rather than scattering volume.}  Indeed,
the strict $V(r)$ gives rise to a completely $k$ independent
scattering amplitude \cite{Baranov,YouPRL}, as is the $k_0 \rightarrow
+\infty$ limit of Eq.~(\ref{symmetry factor}).

Now with $\varphi_{k}$ given by Eq.~(\ref{symmetry factor}) for the
DDI, Eqs. (\ref{eq:PG}), (\ref{eq: Gap Equation}) and
(\ref{eq:number}) form a closed set, which can be solved
self-consistently for $T_c$ as a function of the $p$-wave pairing
strength, $g=-24 \pi D/(5m)$, and for gaps below $T_c$ as a function
of $T$, where $D=md^{2}/2$ is the dipole length.
The unitary limit corresponds to the critical coupling strength
$g_c=-18\pi/mk_0$, at which the scattering length diverges, 
 and a bound state starts to form,
as determined by the Lippmann-Schwinger equation
\cite{Pethick,TLHo}
$g^{-1}_c=-\sum_{\bf{k}}{|\varphi_{\bf{k}}|^2}/({2\epsilon_{\bf{k}}})$,
with $\epsilon_{\bf{k}} = {\bf{k}}^{2}/(2m)$.  Thus $g/g_c = 4 k_0D/15$.
In our numerical
calculations we take $k_0/k_F=20$, corresponding to a dilute case.

\begin{figure}
\centerline{\includegraphics[width=2.8in,clip]{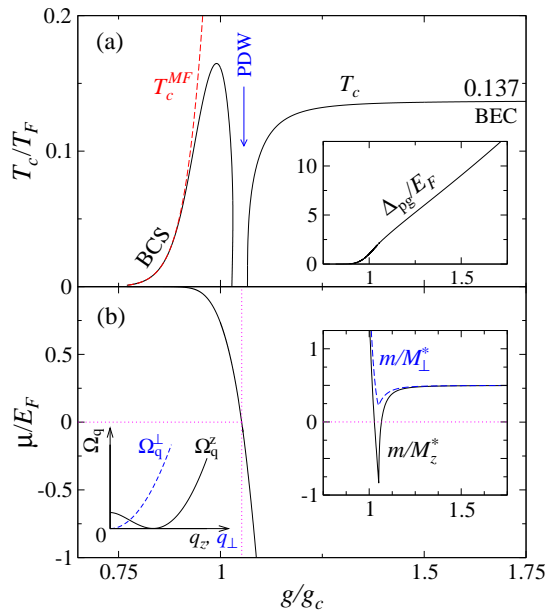}}
\caption{(Color online) (a) Superfluid transition temperature $T_c$
  (black solid curve), the mean-field $T_c^{MF}$ (red dashed curve)
  and (b) chemical potential $\mu(T_c)$ as a function of $g/g_c$.
  Shown in the insets are the pseudogap $\Delta_{pg}(T_c)$ and the
  inverse pair mass $m/M^*$. A PDW state emerges where $T_c$ shuts off
  at intermediate coupling strength and the inverse pair mass
  $m/M^*_z$ becomes negative.  The right insets share the same
  horizontal axis as the main panels. The lower left inset shows
  schematic pair dispersion in the PDW regime. While the inverse mass
  remains positive in the $xy$-plane (blue dashed line), it becomes
  negative in the $z$ direction (black solid curve), with a minimum at
  finite $q_z$ in $\Omega_q^z$.\vspace*{-1.ex}}
\label{fig:phase diagram}
\end{figure}

\vspace*{-1.ex}
\section{Numerical Results and Discussions}
\vspace*{-1.ex}

We first present in Fig.~\ref{fig:phase diagram} the calculated
superfluid transition temperature $T_c$ and corresponding $\mu$ and
pseudogap $\Delta_{pg}$ at $T_c$ as a function of pairing strength,
which are obtained by setting $\Delta_{sc} = 0$.
For comparison, the mean-field solution $T_c^{MF}$ is also shown in
Fig.~\ref{fig:phase diagram}(a) (red dashed curve). In the weak
coupling regime, $T_c$ follows the mean-field BCS result. It starts to
decrease after it reaches a maximum around unitarity $g/g_c=1$, due to
the shrinking Fermi surface. Remarkably, it exhibits a re-entrant
behavior. For a range of intermediate pairing strength, $T_c$ shuts
off completely, before it recovers at stronger couplings, where the
system has entered the BEC regime and all fermions are paired, with
$\mu < 0$. With $M^* $ approaching $2m$ and $n_{pair} = n/2$, $T_c$
approaches the BEC asymptote, 0.137$T_F$, \emph{from below}. The
pseudogap at $T_c$ increases monotonically with $g/g_c$.

In order to understand the re-entrant $T_c$ behavior, we plot the
inverse pair masses in the lower inset of Fig.~\ref{fig:phase
  diagram}(b). It reveals that, when $T_c$ vanishes at the
intermediate pairing strength, the effective pair mass in the dipole
direction, $M^*_z$, at zero momentum becomes negative, so that the
pair dispersion $\Omega_{\bf{q}}$ in the $\hat{z}$-direction becomes
roton-like \cite{rotonGora1,*rotonGora2}, with a minimum at a finite
$q_z$, as shown schematically in the lower left inset of
Fig.~\ref{fig:phase diagram} (solid curve). The pair mass in the
$xy$-plane remains positive. This corresponds to a pair density wave
ground state, with a crystallization wavevector $q_z$ in the
$\hat{z}$-direction. Similar PDW states were extensively investigated in
high $T_c$ superconductors in the quasi-2D context
\cite{TesanovicCPCDW,PDWZhang}.

We emphasize that \emph{the non-monotonic behavior of $T_c$ as a
  function of pairing strength, as found in our $T$-matrix approach of
  the pairing fluctuation theory \cite{ChenPRL}, can be understood on
  physical grounds, without invoking specific details of the theory.}
Indeed, this approach has been accepted by increasingly more
researchers \cite{Torma2,*OzawaBaym,*YangXS,*HeHu,YD05}.  In the weak
coupling regime, $T_c$ follows the mean-field behavior. As the pairing
strength increases towards unitarity, the chemical potential
decreases, leading to a shrinking Fermi surface and thus a decreasing
density of state (DOS) $N(0)\propto \sqrt{\mu}$.  At the same time, a
pseudogap develops gradually due to strong pairing correlations, which
causes a further depletion of the DOS at the Fermi level. Both these
effects cause a reduction of $T_c$, as one can naively expect from
the BCS formula for $T_c$. Such effects will reach their utmost when
the Fermi surface disappears completely at $\mu=0$. Therefore, it is
natural to have a maximum of $T_c$ within the fermionic regime. The
actual position of the maximum depends largely on the range of the
pairing interaction, and is close to unitarity in the contact
potential limit. On the other hand, as the pseudogap develops,
fermions form pairs. Upon entering the bosonic regime, essentially all
fermions are paired. The BEC temperature of these pairs increases with
the pairing strength, as the pair density does. This explains why the
combined $T_c$ exhibits a minimum around $\mu=0$. At this point, the
effective pair mass $M^*$ is significantly heavier than $2m$, due to
the repulsive interaction between pairs.  As the pairing strength
increases further into the BEC regime, the pair size shrinks, and the
inter-pair scattering length decreases, so that $M^*$ decreases
gradually towards $2m$. As a consequence, \emph{the Bose condensation
  temperature $T_c$ of the pairs necessarily increases towards  its
  BEC asymptote from below}. Within a $T$-matrix approximation, these
arguments are independent of the specific form of the pair
susceptibility.

We note that the emergence of the PDW state has to do with the long
range nature of the DDI, which essentially put the system in the high
density regime. At the same time, due to the $p_z$ symmetry, the
coherence length $\xi \sim v_F/\Delta_\mathbf{k}$ diverges in the
nodal $xy$-plane (i.e., $k_z =0$) so that the order parameter
$\Delta_\mathbf{k}=\Delta\varphi_\mathbf{k}$ exhibits a non-local
effect similar to the case of a $d_{x^2-y^2}$-wave superconductor
\cite{Kosztin_nonlocal}. (Here $v_F$ is the Fermi
velocity). \emph{Such a diverging in-plane coherence length makes it
  difficult for the pairs to move in the $\hat{z}$ direction, without
  heavily colliding with other pairs.}  At certain intermediate
interaction strength, pairing is strong while the pair size is large,
so that the repulsive interaction between pairs becomes
strong. Indeed, a careful look at the effective inverse pair mass
reveals that before entering the PDW state, the pair mass already
becomes heavy due to strong pair-pair repulsion. Therefore, the
kinetic energy of the pairs (in the $\hat{z}$ direction) becomes much
smaller than the growing potential energy between pairs, in favor of
forming a Wigner-like crystal structure, which is what we call the PDW
state. Formation of such a crystal structure and minimization of the
pair dispersion at a finite momentum suppress the superfluid $T_c$
down to zero. Such a periodic crystal structure of a PDW state can be
most directly probed using Bragg scattering, similar to the X-ray
diffraction of a crystal structure of a solid.

\begin{figure}
\centerline{\includegraphics[width=3.2in,clip]{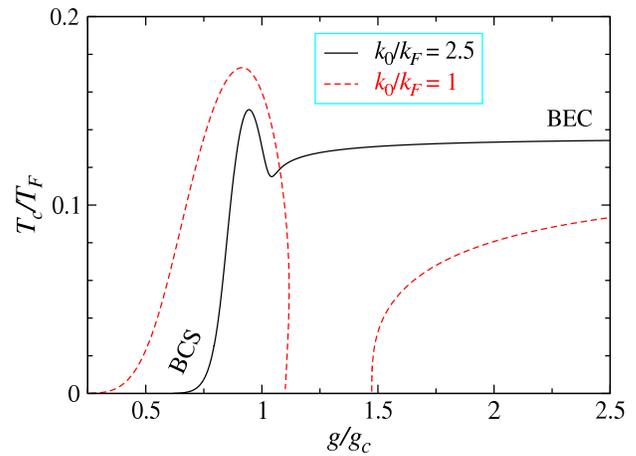}}
\caption{(Color online) $T_c$ behavior of a finite range $p_z^{}$-wave
  superfluid as a function of $g/g_c$ for $k_0/k_F=2.5$ (black solid)
  and 1.0 (red dashed line).  The
  pairing symmetry is given by
  $\left|\varphi_k\right|_p^{}\cos\theta_{\mathbf{k}}$.\vspace*{-1.ex}
}
\label{fig:shortrangepwave}
\end{figure}

To further test this picture, we plotted in
Fig.~\ref{fig:shortrangepwave} the $T_c$ behavior of the finite range
$p_z$-wave superfluid, with a pairing symmetry factor given by
$\left|\varphi_k\right|_p^{}\cos\theta_{\mathbf{k}}$, as a function of
pairing strength for representative values of the range of
interaction, as given by $k_0/k_F = 2.5$ and 1.0. Here $k_F/k_0$
serves as the effective range of interaction, in units of the
interparticle distance ($1/k_F$). For a short range, $k_0/k_F = 2.5$,
the crossover is smooth and continuous, similar to a short range
$s$-wave case \cite{notesonTcmax}, except for a reduced BEC
asymptote. As $k_F/k_0$ increases, more particles are within the range
of interaction at the same time so that the effective repulsion
between pairs becomes strong and the pair mass becomes heavy. For a
larger range, $k_0/k_F = 1$, a reentrant behavior of $T_c$ appears, as
in the dipole-dipole interaction case. (And PDW states emerge where
$T_c$ vanishes). In fact, such reentrant behavior also occurs for
$s$-wave pairing with a large range of interaction
\cite{ChenPhD}. This supports our conclusion that the reentrant
behavior of $T_c$ for a dipolar Fermi gas results from the long range
nature of the DDI. We emphasize that the reentrant behavior is not
unique to the DDI, nor is it to the $p$-wave pairing symmetry.

In the absence of an underlying lattice potential, the PDW state in
the dipolar Fermi gases is distinct from a Mott state. Instead, it may
exhibit behaviors of a Bose metal
\cite{BoseMetal1,*BoseMetal2,Yang_BoseMetal}. The presence of the PDW
manifests a Bose ``surface'' for pair excitations
\cite{Sachdev_BoseSurface,*Paramekanti}, whose energy vanishes at a
finite momentum $q_z$ (with $q_x=q_y=0$). While the pair dispersion
remains positive in the $xy$ plane, the two dimensionality destroys
the long range superfluid order, leading to a metallic ground state
with a density wave of Cooper pairs in the $\hat{z}$ direction. The
nature of the PDW state deserves further systematic investigations
\cite{PDWNote}.

It should be mentioned that the chemical potential $\mu$ changes sign
within the PDW regime. In the fermionic regime, there is a line node
at $k_z=0$ on the Fermi surface in the $p_z$-wave superfluid order
parameter. Once $\mu$ becomes negative, the node disappears and the
excitation spectrum $E_\mathbf{k}$ becomes fully gapped. This may be
regarded as a topological transition \cite{Leggett,Iskin}. The
anisotropy in the pair mass is a consequence of the DDI.  We emphasize
that the re-entrant behavior of $T_c$ is robust against changes of
$k_0$ and independent of the regularization scheme, because $k_0$ does
not modify the long range part of the DDI.  It is also present in the
next leading order, $f_z$-wave channel.

Note that when $\mu$ changes sign, the pairing gap $\Delta$ is rather
large (of the order $E_F$). There exists an extended range of low $T
\ll \Delta$, where $\mu$, $\Delta$ and $M^*$ remain essentially
constant, so that the PDW state is rather insensitive to $T$ in this
temperature range.

For $d$-wave pairing as in the cuprates, $T_c$ vanishes at a lower
critical doping concentration, for which the calculated effective pair
mass diverges as well. Below this doping concentration, the pair
dispersion acquires a minimum at a finite momentum, with a negative
mass at $q=0$. This suggests that the PDW in the cuprates and
the PDW in the dipolar Fermi gases may share the same origin.

\begin{figure}
\centerline{\includegraphics[width=3.2in,clip]{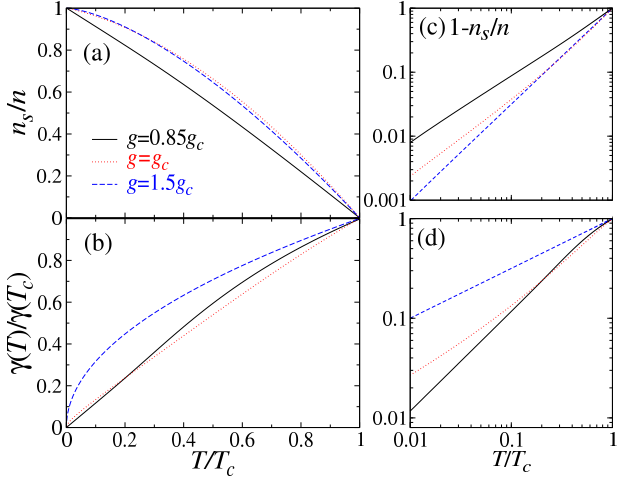}}
\caption{(Color online) Transport and thermodynamic behavior. (a)
  $n_s/n$ and (b) $\gamma(T)/\gamma(T_c)$ as a function of $T/T_c$ for
  $g/g_c=0.85$ (BCS), 1.0 (unitary), and 1.5 (BEC), and log-log plot
  of (c) $1-n_s/n$ and (d) $\gamma(T)/\gamma(T_c)$ vs $T/T_c$.\vspace*{-1.ex}
}
\label{fig:finite temperature}
\end{figure}

Next we investigate the transport and thermodynamics behavior in the
superfluid phase. The superfluid density can be derived using a linear
response theory. Following Ref.~\cite{ChenPRL}, we obtain
\begin{eqnarray}
n_{s}&=&\frac{m\Delta^{2}_{sc}}{3}\sum_{\bf{k}}\frac{1}{E^{2}_{\bf{k}}}\left[\frac{1-2f(E_{\bf{k}})}{2E_{\bf{k}}}
+f'(E_{\bf{k}})\right]\nonumber\\ 
&&{}\times \left[(\nabla_{\bf{k}}\xi_{\bf{k}})^{2}
|\varphi_{\bf{k}}|^{2} - \frac{1}{4}
(\nabla_{\bf{k}}\xi_{\bf{k}}^2)\cdot(\nabla_{\bf{k}}|\varphi_{\bf{k}}|^{2})\right]\,,
\end{eqnarray}
where $f'(x)=\mathrm{d} f(x)/\mathrm{d} x$. It can be shown that
$n_s(0)=n$. At $0 < T \le T_c$, both Bogoliubov quasiparticles 
and pair excitations 
contribute to the thermodynamics. This leads to the
specific heat $C_{v} =
\sum_{\bf{k}}E_{\bf{k}}\partial_{T}f(E_{\bf{k}})+\sum_{\bf{q}}\Omega_{\bf{q}}\partial_{T}b(\Omega_{\bf{q}})$.

Shown in Fig.~\ref{fig:finite temperature} are the $T$ dependencies of
(a) $n_s$ and (b) $\gamma=C_{v}/T$, for $g/g_c=0.85$, 1.0, and 1.5,
corresponding to BCS, unitary and BEC regimes, respectively.  These
two quantities are sensitive to the elementary excitation spectrum.
Due to the line node on the Fermi surface of the $p_z$-wave
superfluid, the low energy density of states $N(E)$ is linear in
$E$. Therefore, the low $T$ superfluid density and specific heat
exhibit power laws in contrast to the exponential behavior of an
$s$-wave superfluid. In the BCS regime, both the low temperature
normal-fluid density $n_{n}/n=1-n_{s} /n$ and $\gamma (T)$ are linear
in $T$, similar to their counterpart in the nodal $d$-wave cuprate
superconductors. On the other hand, in the BEC regime, pair
excitations dominate, so that $n_{n}/n \sim (T/T_c)^{{3}/{2}}$ and
$\gamma \sim (T/T_c)^{{1}/{2}}$, similar to the short-range $s$-wave
case. At $g=g_c$, both types of excitations coexist, and thus the $T$
dependence exhibits a crossover.  The power law behaviors are best
manifested in log-log plots, as slope changes in Fig.~\ref{fig:finite
  temperature}(c) and (d). While the qualitative features shown here
may be easily anticipated, we emphasize that this is the first
systematic study of the thermodynamic behavior of a superfluid of a
dipolar Fermi gas throughout the BCS-BEC crossover.

\begin{figure}
\centerline{\includegraphics[width=3.3in,clip]{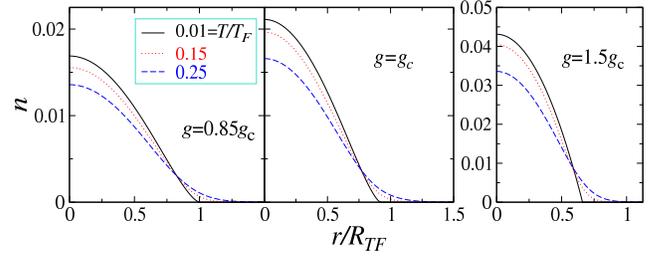}}
\caption{(Color online) Comparison of density profiles in an
  isotropic harmonic trap at $T/T_F=0.01$, 0.15 and 0.25 and pairing
  strengths $g/g_c=0.85$ (BCS), 1.0 (unitary) and 1.5 (BEC). Here
  $R_{TF}$ is the Thomas-Fermi radius and the density $n$ is in units
  of $k^{3}_F$.\vspace*{-1.ex}
}
\label{fig:trap}
\end{figure}

Finally, we consider the effect of a 3D isotropic harmonic trap of
frequency $\omega$ with a trapping potential $V_\text{trap}({\bf{r}})
= \frac{1}{2}m\omega^{2}r^{2}$. We assume that $E_F$ is large enough
to justify the use of LDA \cite{YD05,HeYan}.  Then $\mu$ is replaced
by $\mu ({\bf{r}}) = \mu_{0}-V_\text{trap}({\bf{r}})$, where the
global chemical potential $\mu_0$ is determined by the total fermion
number constraint, $N=\int_\text{trap} n({\bf{r}}) \mathrm{d}^{3}r$,
with local density $n({\bf{r}})$. Outside the superfluid core, a
non-vanishing $\mu_\text{pair}(r)$ is included so that the gap and the
pseudogap equations are extended as $t^{-1}(0, {\bf{0}}) =
Z\mu_\text{pair}$ and $\Delta^2_{pg}=2
Z^{-1}\sum_{\bf{q}}b(\Omega_{\bf{q}}-\mu_\text{pair})$,
respectively. Shown in Fig.~\ref{fig:trap} is the evolution of the
density profile from low to high $T$, throughout the BCS-BEC
crossover. Despite the anisotropic pairing interaction, the density
profile remains isotropic under LDA. It broadens with increasing
temperature whereas it shrinks with increasing DDI strength, similar
to its $s$-wave counterpart with a contact potential \cite{HeYan}. The
isotropic density profile partly reflects the fact that (i) the
pairing symmetry becomes internal degrees of freedom for the fermion
pairs and (ii) within the LDA, this isotropy comes from the isotropic
$V_\text{trap}(r)$. Possible anisotropy in the density profile may occur
when direct pair-pair interactions beyond the $T$-matrix level are
included, without using the LDA.

Recent studies \cite{PuHan,Ronen,SunKai}, using Hartree-Fock
approximation, suggest that the normal state 3D dipolar Fermi gas is
subject to collapse and phase separation instabilities in the high
density and strong DDI regime.  For the dilute case considered in the
present work, the Hartree-Fock contribution to the system energy,
proportional to $n^2$, is relatively weak. Our calculations show that,
within the $T$-matrix approximation, the compressibility for paired
superfluid phase at $T \le T_c$ remains positive definite throughout
the BCS-BEC crossover, ensuring a stable superfluid state. Effects of
direct pair-pair interactions beyond the $T$-matrix approximation will
be investigated in a future study.

\vspace*{-1ex}
\section{Conclusions}
\vspace*{-1.ex}

In summary, our study of single-component dipolar Fermi gases reveals
a re-entrant behavior of a $p_z$-wave superfluid transition $T_c$ and
a PDW state in a range of intermediate DDI strength. Such a PDW state
as well as the $p_z$-wave superfluid phase may be detected using local
density measurements, Bragg spectroscopy and momentum resolved rf
spectroscopy.

\vspace*{-1.ex}
\acknowledgments
\vspace*{-1.5ex}

We thank Hui Zhai, Wei Yi, Xin Wan, Hua Chen and K. Levin for helpful
discussions.  This work is supported by NSF of China (Grants
No. 10974173 and No. 11274267), the National Basic Research Program of
China (Grants No. 2011CB921303 and No. 2012CB927404), and NSF of
Zhejiang Province of China (Grant No.  LZ13A040001).

\bibliographystyle{apsrev} 

\begin{thebibliography}{51}
\bibitem[{Not()}]{NoteonRegularization}
\bibinfo{note}{This regularization is justified in that the actual interaction
  necessarily deviates from the strict DDI at a distance closer than or
  comparable to the size of the atoms (or molecules).}
\bibitem[{not()}]{notesonTcmax}
\bibinfo{note}{Note that the location of the maximum $T_c$ depends on
  $k_0/k_F$. Only in the $k_0 \rightarrow +\infty$ limit does the maximum
  occurs at $g/g_c=1$.}
\bibitem[{PDW()}]{PDWNote}
\bibinfo{note}{The zero $T$ inverse pair mass in the PDW state is extracted
  from the pairing $T$ matrix, based on the mean-field solution of the gap
  $\Delta$ and chemical potential $\mu$, since it is generally believed that
  the mean-field result is reasonably good for intermediate pairing strengths
  at $T=0$. It remains to see whether such a PDW state is present in the $GG$
  scheme of the $T$-matrix approximation, with pair susceptibility $\chi =
  GG$.}
\end{thebibliography}

\end{document}